# On the Evolution and Gravitational Collapse of a Toroidal Vortex

K.Yu. Bliokh and V.M. Kontorovich

*Institute of Radio Astronomy, 4 Krasnoznamyonnaya st., Kharkov, 61002, Ukraine*
*e-mail: vkont@ira.kharkov.ua*

The evolution and collapse of a gaseous toroidal vortex under the action of self-gravitation are considered using the Hamiltonian mechanics approach. It is shown that evolution occurs in three main stages separated by characteristic time scales. First, a compression along the small radius to a quasi-equilibrium state takes place, followed by a slower compression along the large radius to a more stable compact vortex object. In the latter stage, the possibility of effective scattering and ejection of particles along the vortex axis (jet formation) is detected. As a result, mass, energy, and momentum losses take place, and the vortex collapses.

## 1. Introduction

Vortices, which are traditionally described in the framework of incompressible fluid dynamics [1,2], are special objects of application of Hamiltonian methods. In the epoch of solitons, the interest in vortices as allied localized formations has grown considerably. The current state of this problem is described in reviews and articles [3–7] and in the references cited therein.

Compressibility and self-gravitation of such objects may play a significant role in astrophysical applications [8–10]. In some cases (e.g., in the vicinity of compact objects at the centers of galaxies), the effect of the external gravitational field on these objects may be significant. Both these possibilities will be taken into account in this study.

Another extremely important problem in contemporary astrophysics is associated with the origin of cosmic jets arising, according to prevailing concepts, in accretion disks of various origins and scales (from galactic [11] to stellar [12]). The interpretation of such jets is a nontrivial problem. In spite of considerable advances made in this direction in the framework of magnetohydrodynamics (MHD) [13], serious difficulties still remain since solutions were obtained in a special geometry and strong magnetic fields are required. It will be shown below that the formation of (unidirectional) jets is a natural consequence of the evolution of self-gravitating vortices and is an indispensable condition of their collapse. Jets are generated in zero magnetic fields also.

A possible relation between this problem and the existence of occluding toroids[1] in the vicinity of central compact objects in the active galactic nuclei [14].

## 2. First stage: evolution of a thin vortex

We assume that the shape of a vortex at the initial stage is a thin toroid (Fig. 1) whose radii satisfy the inequality

$$r \ll R . \qquad (1)$$

---

[1] We do not touch upon the classical problem of stability of toroids rotating as a single entity, which stems to Poincaré and Dyson. The modern state of affairs in the framework of the general theory of relativity and references can be found in [15].

(An example of such a vortex in fluid dynamics is a Maxwellian vortex [1,2].) We will henceforth assume that the vortex evolution at the initial stage occurs without a change in the toroidal shape. Thus, in addition to a rotational degree of freedom, the system possesses two translational degrees of freedom corresponding to changes in $r$ and $R$ (we disregard the translational motion of the vortex as a whole). It should be noted that, in view of condition (1), the motion in $r$ and rotation constitute a local compression of a rotating cylinder, while the motion in $R$ indicates the collapse of a thin ting.

We write the system Hamiltonian in the form

$$H = \frac{1}{2M}\left[p_r^2 + p_R^2 + \frac{p_\varphi^2}{r^2}\right] + U(r,R) . \quad (2)$$

Here, $M$ is the total mass of the vortex, $p_s$ are the momenta corresponding to coordinates $s$, $\varphi$ is the cyclic coordinate of rotation, and $U$ is the gravitational potential energy of the system. Hamilton equations corresponding to Hamiltonian (2) have the form

$$\dot{r} = \frac{p_r}{M}, \quad \dot{R} = \frac{p_R}{M}, \quad \dot{\varphi} = \frac{p_\varphi}{Mr^2} ;$$

$$\dot{p}_r = \frac{p_\varphi^2}{Mr^3} - \frac{\partial U}{\partial r}, \quad \dot{p}_R = \frac{\partial U}{\partial R}, \quad \dot{p}_\varphi = 0. \quad (3)$$

This leads to the equations of motion for the translational degrees of freedom,

$$\ddot{r} = \frac{p_\varphi^2}{M^2 r^3} - \frac{1}{M}\frac{\partial U}{\partial r} ,$$

$$\ddot{R} = -\frac{1}{M}\frac{\partial U}{\partial R} , \quad (4)$$

and to the integral of motion for the rotational degree of freedom:

$$p_\varphi = Mr^2\dot{\varphi} = const . \quad (5)$$

This integral of motion expresses the angular momentum conservation law (the quantity $2\pi p_\varphi / M$ in fluid dynamics corresponds to vorticity).

Let us now define function $U(r,R)$. We consider first Eq. (4a) describing the evolution of a rotating cylinder of radius $r$. The gravitational force acting on a test particle on the surface of the cylinder is given by

$$F_r = -G\frac{2\chi m}{r} , \quad (6)$$

where $G$ is the gravitational constant, $\chi$ is the mass of the cylinder per unit length, and $m$ is the mass of the test particle. Thus, the gravitational force appearing on the right-hand side of Eq. (4a) has the form

$$-\frac{1}{M}\frac{\partial U}{\partial r} = -G\frac{2\chi}{r} . \quad (7)$$

In the present case of a thin toroid, we have $\chi = M/2\pi R$, whence

$$U(r,R) = G\frac{M^2}{\pi R}\ln\frac{r}{R} + c_1(R) . \quad (8)$$

In order to determine the dependence of potential energy $U$ on large radius $R$, we consider the second equation of motion, Eq. (4b). It must describe the gravitational contraction of a thin ring of radius $R$. The force acting on a test particle located on an infinitely thin ring is given by

$$F_R = -G\frac{Mm}{2\pi R^2}\int_0^\pi \frac{d(\vartheta/2)}{\sin(\vartheta/2)} = \infty . \quad (9)$$

This formula can be derived by direct integration of the contributions from the interaction of the particle will all elements of the ring. In order to avoid divergence for $\vartheta \to 0$, we must take into account the finite thickness of the ring. For this purpose, we truncate the diverging part of expression (9), replacing the integration domain $(0,\pi)$ by $(\vartheta_c,\pi)$, where $\vartheta_c = \alpha r/R$ ($\alpha \sim 1$ is a numerical factor). This gives

$$U(r,R) = G\frac{M^2}{2\pi R}\ln\frac{\alpha r}{R} - G\frac{M^2}{2\pi R} + c_2(r) \ . \tag{10}$$

where $\tilde{\alpha} = \alpha/e$. Setting $c_2 = 0$ in Eq. (10) and $c_1(R) = GM^2 \ln\tilde{\alpha}/\pi R$ in Eq. (8), we note that formulas (8) and (10) can be reduced to the same form (the difference will be only in coefficient 1/2). Such a difference is insignificant for our analysis, and we assume the true numerical coefficient in formula (8).

We can now write the Hamiltonian (2) of a thin toroidal vortex:

$$H = \frac{1}{2M}\left[p_r^2 + p_R^2 + \frac{p_\varphi^2}{r^2}\right] + G\frac{M^2}{\pi R}\ln\frac{\tilde{\alpha} r}{R} \ . \tag{11}$$

In this case, the equations of motion (4) assume the form

$$\ddot{r} = \frac{p_\varphi^2}{M^2 r^3} - G\frac{M}{\pi R r} \ , \tag{12a}$$

$$\ddot{R} = -G\frac{M}{\pi R^2}\ln\frac{\alpha r}{R} \ . \tag{12b}$$

In view of condition (1), we have

$$\frac{r}{R}\ln\frac{r}{R} \ll 1 \ .$$

Than the force of gravitational attraction along $r$ (12a) is much stronger than the force of gravitational contraction along $R$ (12b). This allows us to divide the evolution of the system in different scales: fast (in $r$) and slow (in $R$). It is natural to assume that fast evolution (for practically constant $R$) leads to the establishment of equilibrium in Eq. (12a). In this case, the force of gravitational attraction is compensated by the centrifugal rotational force:

$$\frac{p_\varphi^2}{M^2 r} - G\frac{M}{\pi R r} = 0 \ . \tag{13}$$

whence

$$r = \sqrt{\frac{\pi p_\varphi^2 R}{GM^3}} \ .$$

This expression determines the equilibrium small radius as a function of the large radius, $r_{eq} = r(R)$, while the inequality $r > r_{eq}$ corresponds to the criterion of gravitational instability with the Jeans scale $r_{eq}$. The latter becomes obvious if we assume that the toroid mass is $M = \pi r^2 R \rho$, where $\rho$ is its density [8–10].

As a rough estimate of the time of vortex contraction along small radius $r$ to the quasi-equilibrium state, we can use the expression

$$t_1 \simeq \sqrt{\left|\frac{r_0 - r_{eq}(R_0)}{F_r(r_0, R_0)}\right|} = \sqrt{\left|\frac{M^2 R_0^4 - p_\varphi r_0^3 \sqrt{\pi M R_0/G}}{p_\varphi^2 - GM^3 r_0^2/\pi R_0}\right|} \ , \tag{14}$$

where $r_0$, $R_0$, and $F_r$ are the initial values of the small and large radii of the vortex and the force appearing on the right-hand side of Eq. (12a), respectively.

If inequality (1) holds at the initial instant, the toroidal vortex will experience, in accordance with the equation of motion (12b) and relation (13), a slow contraction along both radii until they become on the same order of magnitude:

$$r \sim R = R_c = \frac{\pi p_\varphi^2}{GM^3} = \frac{r_{eq}^2}{R} \ll r_0 \ . \tag{15}$$

At this stage, the initial assumptions (1) are violated, and the description used above becomes inapplicable. It is impossible in this case to divide the vortex evolution in two translational degrees of freedom $r$ and $R$, and the vortex should be described as a single compact object with a complex internal structure.

Let us determine some of the most important parameters of the vortex, characterizing it at the second stage (15). Using the angular momentum conservation law (5) and (13), we obtain the following expression for the velocity of particles on the vortex surface:

$$v = \frac{p_\varphi}{Mr} \sim \sqrt{\frac{GM}{\pi R}} \sim \frac{GM^2}{\pi p_\varphi} \frac{r}{R} \ . \tag{16}$$

It can be seen that the rotational velocity of a vortex, which is in equilibrium in $r$, is determined only by its mass and the large radius and increases upon contraction along $R$.[2]

When the vortex reaches the end of the first stage, the velocity attains the value

$$v_{fin} \sim \frac{GM^2}{\pi p_\varphi} \ . \tag{17}$$

and becomes much larger than the initial velocity $v_0$:

$$\frac{v_{fin}}{v_0} \sim \sqrt{\frac{R_{fin}}{R_c}} \sim \frac{r_{fin}}{R_c} \sim \frac{R_{fin}}{r_0} \gg 1 \ , \tag{18}$$

In analogy with relation (14), we can estimate the time of the vortex contraction along the large radius as

$$t_2 \sim \sqrt{\frac{R_0 - R_c}{F_R[r_{eq}(R_0), R_0]}} = \sqrt{\frac{\pi R_0^2 (R_0 - \pi p_\varphi^2 / GM^3)}{GM \ln(\alpha \sqrt{\pi p_\varphi^2 / GM^3 R_0})}} \sim t_1 \tag{19}$$

where $F_R$ is the force appearing on the right-hand side of Eq. (12b).

Let us also consider the distribution of the kinetic energy acquired by the vortex over the degrees of freedom. We assume that the substance in the vortex is initially almost free and its potential energy and kinetic energy (11) are close to zero. Then, at the critical stage (15) of the collapse, the potential and kinetic energies of the substance can be estimated as

$$U_{fin} \sim -\frac{GM^2}{\pi R_c} \ln \tilde{\alpha} \sim -\frac{GM^2}{\pi R_c} = -\frac{G^2 M^3}{\pi^2 p_\varphi^2} \ . \tag{20}$$

$$T_{fin}^{(rot)} \sim \frac{p_\varphi^2}{2MR_c^2} = \frac{G^2 M^5}{2\pi^2 p_\varphi^2} \ .$$

(It should be noted that this result is in accordance with the virial theorem for $U \sim R^{-1}$; i.e. $E = -T < 0$.) Since the total energy is an integral of motion and its initial energy is close to zero, relations (11) and (20) imply that the kinetic energy of translational motion is of the same order of magnitude. Thus, we can conclude that the kinetic energy released during the collapse is distributed uniformly (in order of magnitude) between the rotational and translational degrees of freedom. This fact will be important for the subsequent analysis; in particular, this means that, if the rotational velocity of a particle of the substance is less than doubled at random, it is sufficient for the particle detachment and escape from the system.

---

[2] It should be noted that this is due to the fact that momentum $p_\varphi$ is cancelled out in relation (16). This is a consequence of the equilibrium condition (13), whose form is determined in turn by the form of potential (11).

## 3. Second stage: evolution of a compact vortex

Let us try to imagine the scenario of contraction of a gravitating vortex, when it is a compact object topologically equivalent to a toroid. We can expect that, under the action of gravitational forces, it will approach a certain spheroidal configuration resembling a Hill vortex [1,2] (Fig. 2).

If we consider such an object as an estimate, we can assume that it possesses a rotational and a translational degree of freedom. The latter is determined by a change in its radius $R$. The rotational radius of particles (which was the independent quantity $r$ in the previous section) is now approximately equal to $R/2$. The Hamiltonian of such a vortex can be written in the form

$$H = \frac{p_R^2}{2M} + \frac{2p_\varphi^2}{MR^2} - \frac{GM^2}{R} \quad . \tag{21}$$

Here, we have assumed that the form of the second term is the same as in relation (11) with $r = R/2$ and had taken the potential energy of a sphere for the potential energy of the vortex. Hamiltonian (21) corresponds to the equation of motion

$$\ddot{R} = \frac{4p_\varphi^2}{M^2 R^3} - \frac{GM}{R^2} \quad . \tag{22}$$

This equation has an equilibrium position, when

$$\frac{4p_\varphi^2}{M^2 R^3} - \frac{GM}{R^2} = 0 \text{ , or } R_{c1} = \frac{4p_\varphi^2}{GM^3} \quad , \tag{23}$$

where $R_{c1}$ plays the role of the Jeans scale as before. Thus, this object is in equilibrium for a radius on the same order of magnitude as that at which the first stage of evolution terminates (cf. relation (15)). This means that, in the problem on the collapse of a thin toroidal vortex, there is no need to consider the evolution of a compact vortex separately. We can assume that equilibrium sets in immediately after the ring acquires parameters (15)–(18).

On the other hand, we can consider the problem of contraction of a vortex, which resembles a Hill vortex from the very outset (Fig. 2), but is initially far from equilibrium. Then we ultimately arrive at an equilibrium compact vortex with a radius on the same order of magnitude as the radius in (23). The rotational velocity of particles in this case is of the order of velocity (16):

$$v_{fin} \sim \frac{GM^2}{2p_\varphi} \quad , \tag{24}$$

where

$$\frac{v_{fin}}{v_{in}} \sim \frac{R_{in}}{R_{c1}} \gg 1 .$$

Similarly, it can be easily proved that, upon the establishment of equilibrium (23), at least half the released potential energy is transformed into the kinetic energy of rotation (the remaining part being transformed into heat).

## 4. Scattering and detachment of particles

Thus, after various possible stages of evolution, a toroidal vortex is transformed into a compact object with characteristic parameters (15), (17), and (18) (or (23) and (24)) (see Fig. 2). The rotation velocity of the substance in it is much higher than in the initial vortex. It is worth noting that flows of matter passing through the vortex in the vicinity of its axis are closely spaced. This means that effective scattering of particles may take place in this region. Such a scattering will obviously increase the velocity of a certain fraction of particles by a factor of $\delta \sim 1$. In accordance with the arguments given at the end of the previous two sections, a less than double increase in the rotational velocity of particles is enough for gathering a kinetic energy sufficient for detachment. Consequently, we can expect that a certain fraction of particles from the flows passing along the axis of a compact vortex acquire a sufficient. energy as a result of scattering and are ejected from the vortex. Thus, a directional jet carrying away the matter from the center of the vortex can be formed. (Here, we disregard the change in the vortex configuration that may take place as a result of the mass loss due to such ejection.)

Let us consider one more argument illustrating the above scenario. In the Appendix, we will consider the motion of a test particle in the gravitational field of a ring (thin toroid) with a fixed radius (Fig. 3). For low energies, the particle rotates in a small-radius orbit wound around the ring (Fig. 3a). This motion corresponds to a thin vortex (the possible first stage of the evolution). As the particle energy increases, various complex trajectories appear; however, the orientations of these trajectories do not correspond to vortex motion and we will not consider such trajectories here. Finally, starting from a certain energy value, the particle passes to almost closed trajectories of a figure-of-eight shape (Fig. 3d). The rotational radius of particles becomes on the order of the ring radius (15), which corresponds precisely to the final sage of vortex contraction. The kinetic energy of a particle in such trajectories is close to the energy required for the detachment of particles. The motion of particles in "figure-of-eights" will lead to their effective collisions and scattering in the vicinity of the vortex axis.[3]

We can state that the toroid contraction has qualitatively the same consequences for moving particles as an increase in their energy for a fixed size of the toroid. Obviously, a tendency ultimately leading to the detachment of a fraction of particles exists, the most favorable conditions for this effect being created in the vicinity of the vortex axis. In the long run, this leads to the emergence of an axial (unilateral) jet carrying away the energy, mass, and angular momentum of the vortex. As a result, the vortex contraction will continue (resulting in collapse), the contraction rate $dR/dt$ being determined by the vortex mass loss rate (particle flux in the jet; see below). Thus, the vortex collapse and the emergence of a jet are correlated unambiguously.

## 5. Vortex collapse

Let us consider the consequences of the ejection of particles from a vortex according to the scenario proposed in the previous section. The particle flow carries away the mass, energy, and angular momentum of the vortex. The latter quantities can be estimated as

$$E \sim -\frac{mv^2}{2} \ , \quad p_\varphi \sim \frac{MRv}{2} \ . \qquad (25)$$

where all the quantities correspond to an equilibrium compact vortex (see Section 3) and $E = -T$ (see above). Relations (25) lead to

$$R \sim \frac{p_\varphi}{\sqrt{-ME}} \ . \qquad (26)$$

Differentiating this relation with respect to time, we obtain

$$\dot R \sim \frac{1}{\sqrt{-ME}}\dot p_\varphi - \frac{p_\varphi}{2(-ME)^{3/2}}\left(M\dot E + E\dot M\right) , \qquad (27)$$

We assume that the particle flux is comparatively small and this process occurs at a much lower rate than the rate of establishment of equilibrium of the compact vortex. In this case, we can estimate the change in the characteristics of the vortex carried away by the particle flow as

$$\dot M \sim -J \ , \quad \dot E \sim J\frac{v^2}{2} \sim -\frac{J}{M}E \ , \quad \dot p_\varphi \sim -J\frac{Rv}{2} \sim -\frac{J}{M}p_\varphi \ , \qquad (28)$$

---

[3] The existence of flows of matter of the figure-of-eight type also follows from the hydrodynamic model of a Maxwell vortex (see, for example, [2]).

where $J$ is the mass flux in the ejected jet of matter. Substituting relations (28) into (27) and taking into account relation (26), we obtain

$$\dot{R} \sim -\frac{J}{M}R = -\frac{\dot{M}}{M}R \; . \qquad (29)$$

The solution to this equation has the form

$$R(t) = R(0)\left[\frac{M(t)}{M(0)}\right]^{-\beta} . \qquad (30)$$

Here, $\beta \sim 1$ is a certain positive constant (emerging due to the fact that we obtained above only the order-of-magnitude estimates for the vortex parameters), and the initial instant of time corresponds to the arrival of the vortex at the compact equilibrium state and to the beginning of the effective scattering and detachment of particles. The time dependence of the vortex mass $M(t)$ is determined for the specific mechanism of particle scattering. In the general case, the mass flux of matter, $J = -\dot{M}$, is a function of the main vortex parameters: mass, energy, and angular momentum. If we assume in the simplest case that the flux of matter is proportional to the vortex mass and weakly depends on other parameters ($J = kM$), Eqs. (28)–(30) will lead to the exponential laws

$$M(t) = M(0)\exp(-kt) \; , \quad R(t) = R(0)\exp(-\beta kt) \; . \qquad (31)$$

Thus, Eqs. (29)–(31) show that the scattering of particles and ejection of matter indeed lead to the collapse of a compact vortex.

The characteristic time scale of the collapse is defined as

$$t_{col} \sim \left|\frac{M}{\dot{M}}\right| . \qquad (32)$$

In accordance with the above assumption, the collapse must be slow as compared to the characteristic time of the vortex contraction to the equilibrium state, which corresponds to $t_{col} \sim t_2$.

## 6. Generalizations

We can easily generalize the above analysis to the case when a system contains a massive body at its center and when the vortex rotates about its axis. These factors lead to the emergence of additional terms in Eq. (12):

$$\ddot{R} = -G\frac{M}{\pi R^2}\ln\frac{\alpha r}{R} - G\frac{\widetilde{M}}{R^2} + \frac{p_\vartheta^2}{M^2 R^3} \; . \qquad (33)$$

Here, $M$ is the mass of the central object and $p_\vartheta = MR^2\dot{\vartheta} = const$ is the angular momentum associated with the rotation of the toroid about its axis. The supplementary terms do not affect in any way the evolution of the system along the small radius. Their effect on the vortex evolution along $R$ can be divided into the following limiting cases.

1. If $\max\left(\frac{G\widetilde{M}}{R_c^2}, \frac{p_\vartheta^2}{M^2 R_c^3}\right) \ll \frac{GM}{\pi R_c^2}$, the effect of these terms can be disregarded, and the entire dynamic analysis carried out in Sections 2 and 3 as well as the corresponding conclusions remain in force. However, the presence of the central mass in the region of the most probable intersection of particle flows may affect their scattering and detachment.

2. If $\frac{\pi\widetilde{M}}{M} \ll 1$, and $\frac{\pi p_\vartheta^2}{GM^3 R_c} \gg 1$, the rotation of the toroid about its axis arrests contraction before it reaches its critical stage $r \sim R \sim R_c$. The equilibrium state corresponds to the large radius defined by the relation

$$\frac{GM^3 R}{\pi p_\vartheta^2} \ln \alpha \sqrt{\frac{\pi p_\varphi^2}{GM^3 R}} = 1 ,$$

and to the small radius defined by substituting the large radius into relation (18). In this case, the probability of effective scattering and detachment of particles at the middle of the vortex virtually vanishes and, hence, collapse does not take place.

3. If $\frac{\pi \widetilde{M}}{M} \gg 1$ and $\frac{\pi p_\vartheta^2}{GM^3 R_c} \ll 1$, the revolution of a vortex around its axis is insignificant, and the central mass enhances the contraction. The vortex contracts to a compact object and its subsequent behavior depends on the scenario of direct interaction of the vertex with the central mass. Naturally, the scattering of particles and the possible vortex collapse in this case also depend to a considerable extent on the interaction of the matter with the central mass.

4. If $\frac{\pi \widetilde{M}}{M} \gg 1$ and $\frac{\pi p_\vartheta^2}{GM^3 R_c} \gg 1$, the last two terms on the right-hand side of Eq. (33) compete. If the first term is greater than the second (the attraction of the central object prevails), the situation corresponds to case 3; for the opposite relation, we have case 2.

## Acknowledgments

This study was partly supported by the INTAS (grant no. 00-00292).

## Appendix: motion of particles in the gravitational field of the ring

We assume that the gravitational field of a thin toroid is close to the field of an infinitely thin ring of the same mass. Let the ring radius be $r_0$ and $r$, $\varphi$ and $z$ be the cylindrical coordinates, $z = 0$ corresponding to the plane of the ring. The gravitational potential of the ring is given by

$$U(r,z) = -\frac{GM}{2\pi} \int_0^{2\pi} \frac{d\varphi}{\sqrt{z^2 + r^2 + r_0^2 - 2rr_0 \cos\varphi}} .$$

Introducing the dimensionless variables $\xi = z/r_0$ and $\rho = r/r_0$ and time $\tau = t\sqrt{GM/2\pi r_0^3}$ we obtain the equations of motion of a test particle in this potential:

$$\rho'' = -\int_0^{2\pi} \frac{(\rho - \cos\varphi)}{(\xi^2 + \rho^2 + 1 - 2\rho \cos\varphi)^{3/2}} d\varphi ,$$

$$\xi'' = -\int_0^{2\pi} \frac{\xi}{(\xi^2 + \rho^2 + 1 - 2\rho \cos\varphi)^{3/2}} d\varphi ,$$

where the primes indicate differentiation with respect to $\tau$, and we assume that $\varphi = 0$ for a particle.

Figure 3 shows typical results of numerical calculations based on these equations for a finite motion. The trajectories lie in the $(\rho, \xi)$ plane and are given in the increasing order of the particle energy. It can easily be seen that, in trajectories of the "dove-tail" type (see Figs. 3b and 3e), a particle moves practically along the same curve in opposite directions; consequently, such motion cannot be maintained in the framework of collective motion of particles since the latter motion would inevitably lead to collisions and strong scattering. In addition, the trajectories in Fig. 3c and 3f cannot exist for a collective vortex motion of particles since different segments of a trajectory correspond to opposite directions of vorticity. Thus, only the trajectories in Figs. 3a and 3d can exist in the framework of collective vortex motion of particles. The trajectory in Fig. 3a is the cross section of a thin toroidal vortex of the Maxwellian vortex type considered in Section 2, while the "figure-of-eight" in Fig. 3d can appear during motion of particles in a compact vortex of the type of a Hill vortex emerging at the late stage of contraction (see Section 3).

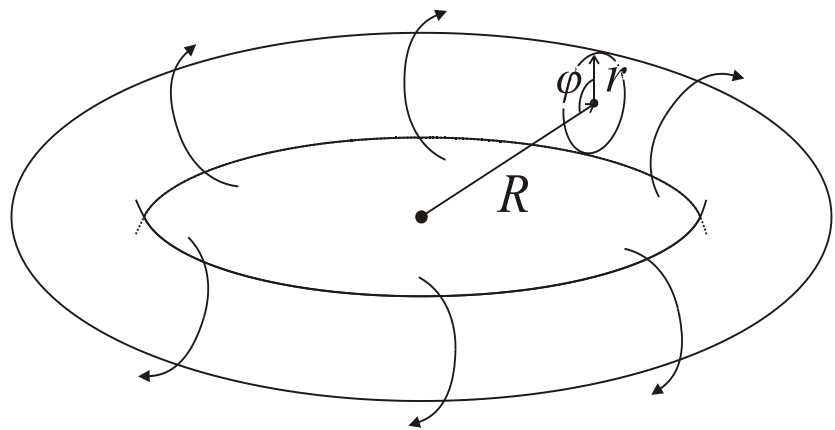

**Fig. 1.** Thin toroidal vortex of the Maxwellian type.

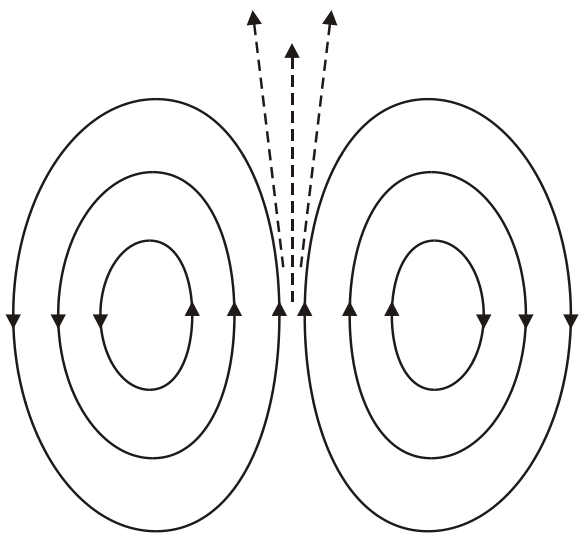

**Fig. 2.** Cross section of a compact spheroidal vortex of the Hill vortex type.

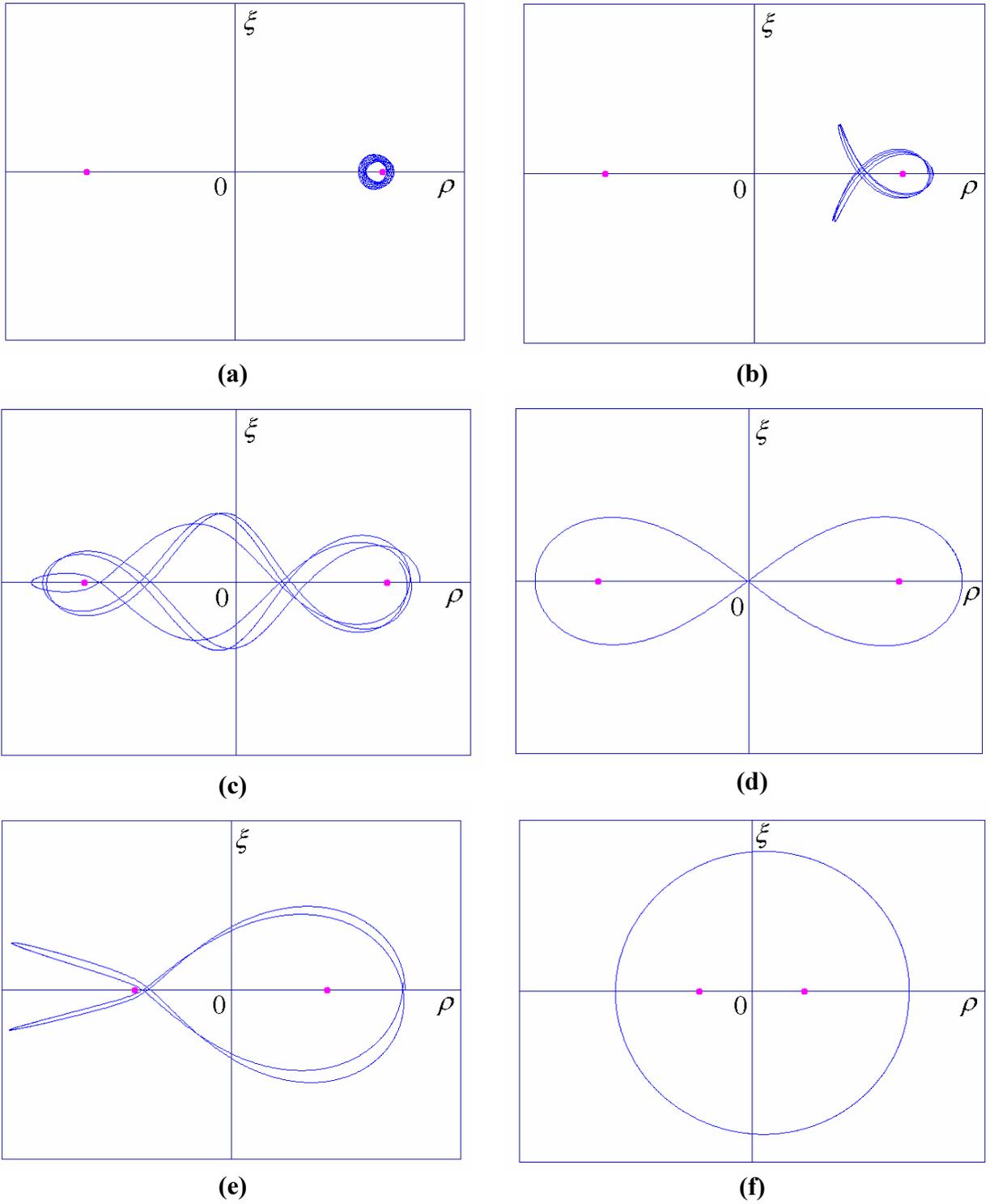

**Fig. 3.** Typical finite trajectories of a particle in the gravitational field of the ring. In the plane $(\rho = r/r_0, \xi = z/r_0)$, the particle moves around two attracting centers formed as a result of the dissection of the ring. The initial conditions are $\rho = 1 + \rho_0$, $\xi = 0$, $\dot\rho = 0$, and $\dot\xi = \sqrt{2}$ (the value of $\dot\xi$ is chosen to coincide with the "orbital velocity" in Eq. (12a), which is valid in the vicinity of attracting centers). (a) $\rho_0 = 0.08$, rotation around a single center; (b) $\rho_0 = 0.17$, "dove-tail" type motion around a single center; (c) $\rho_0 = 0.22$, motion of the double "figure-of-eight" type around two centers; (d) $\rho_0 = 0.42$, motion of the "figure-of-eight" type around two centers; (e) $\rho_0 = 0.81$, motion of the "dove-tail" type around two centers; (f) $\rho_0 = 2$, rotation around two centers.